\documentclass[twocolumn]{aastex63}

\received{\textit{2025 December 29}}
\revised{\ldots}
\accepted{\ldots}
\published{\ldots}

\shorttitle{Secular Variation}
\shortauthors{Liu et al. 2026}

\usepackage{graphicx,times}
\usepackage{natbib}
\usepackage{amssymb,amsmath}
\usepackage{xurl}

\usepackage{threeparttable}

\def\fermi {\emph{Fermi}}
\def\xmm {\emph{XMM-Newton}}
\def\Star{PSR~J2021$+$4026}

\begin{document}

\title{Secular Evolution of {\Star}: Long-Term $\gamma$-Ray Flux and Spin-Down Variability Beyond State Transitions}

   \author[0009-0008-3286-7254]{Xue-Zhi Liu}
   \affil{Laboratory for Compact  Object Astrophysics and AstroSpace Technology, Central China Normal University, Wuhan 430079, People's Republic of China; \url{xz\_liu@mails.ccnu.edu.cn}; \url{zhxp@ccnu.edu.cn}}
   \affil{Institute of Astrophysics, Central China Normal University, Wuhan 430079, People's Republic of China}
   \affil{State Key Laboratory of Particle Astrophysics, Institute of High Energy Physics, Chinese Academy of Sciences, Beijing 100049, People's Republic of China; \url{gemy@ihep.ac.cn}}
   
   \author[0000-0002-2749-6638]{Ming-Yu Ge}
   \affil{State Key Laboratory of Particle Astrophysics, Institute of High Energy Physics, Chinese Academy of Sciences, Beijing 100049, People's Republic of China; \url{gemy@ihep.ac.cn}}
   \affiliation{University of Chinese Academy of Sciences, Chinese Academy of Sciences, Beijing 100049, People's Republic of China}

   \author[0000-0001-8868-4619]{Xiao-Ping Zheng}
   \affil{Laboratory for Compact  Object Astrophysics and AstroSpace Technology, Central China Normal University, Wuhan 430079, People's Republic of China; \url{xz\_liu@mails.ccnu.edu.cn}; \url{zhxp@ccnu.edu.cn}}
   \affil{Institute of Astrophysics, Central China Normal University, Wuhan 430079, People's Republic of China}

   \author[0000-0003-4585-589X]{Xiao-Bo Li}
   \affil{State Key Laboratory of Particle Astrophysics, Institute of High Energy Physics, Chinese Academy of Sciences, Beijing 100049, People's Republic of China; \url{gemy@ihep.ac.cn}}
    \affiliation{University of Chinese Academy of Sciences, Chinese Academy of Sciences, Beijing 100049, People's Republic of China}
  
   \author[0009-0009-8477-8744]{Han-Long Peng}
   \affil{School of Physics and Technology, Nanjing Normal University, Nanjing, 210023, Jiangsu, People's Republic of China}

   \author[0000-0002-1662-7735]{Wen-Tao Ye}
   \affil{State Key Laboratory of Particle Astrophysics, Institute of High Energy Physics, Chinese Academy of Sciences, Beijing 100049, People's Republic of China; \url{gemy@ihep.ac.cn}}
   \affiliation{University of Chinese Academy of Sciences, Chinese Academy of Sciences, Beijing 100049, People's Republic of China}

   \author[0009-0008-0518-6795]{Bo-Yan Chen}
   \affil{State Key Laboratory of Particle Astrophysics, Institute of High Energy Physics, Chinese Academy of Sciences, Beijing 100049, People's Republic of China; \url{gemy@ihep.ac.cn}}
   \affiliation{University of Chinese Academy of Sciences, Chinese Academy of Sciences, Beijing 100049, People's Republic of China}

   \author[0000-0003-2256-6286]{Shi-Jie Zheng}
   \affil{State Key Laboratory of Particle Astrophysics, Institute of High Energy Physics, Chinese Academy of Sciences, Beijing 100049, People's Republic of China; \url{gemy@ihep.ac.cn}}
   \affiliation{University of Chinese Academy of Sciences, Chinese Academy of Sciences, Beijing 100049, People's Republic of China}   
   \author[0000-0003-3248-6087]{Fang-Jun Lu}
   \affil{State Key Laboratory of Particle Astrophysics, Institute of High Energy Physics, Chinese Academy of Sciences, Beijing 100049, People's Republic of China; \url{gemy@ihep.ac.cn}}
   \author[0000-0001-5586-1017]{Shuang-Nan Zhang}
   \affil{State Key Laboratory of Particle Astrophysics, Institute of High Energy Physics, Chinese Academy of Sciences, Beijing 100049, People's Republic of China; \url{gemy@ihep.ac.cn}}
   \affiliation{University of Chinese Academy of Sciences, Chinese Academy of Sciences, Beijing 100049, People's Republic of China}

\begin{abstract}
   {\Star} is a remarkable $\gamma$-ray pulsar exhibiting repeated transitions between high $\gamma$-ray flux (HGF) and low $\gamma$-ray flux (LGF) states. With 17-yr {\fermi}-LAT monitoring, we reveal persistent secular evolution and enhanced spin-down rate variability \emph{within} individual emission states---beneath the quasi-periodic state transitions. After removing discrete jumps, the jump-corrected flux $\delta F_\gamma$ shows a three-phase evolution: rise ($+2.02^{+0.17}_{-0.15}\%~\mathrm{yr}^{-1}$), decline ($-3.72^{+0.34}_{-0.47}\%~\mathrm{yr}^{-1}$), and rapid rise ($+14.9^{+6.4}_{-4.4}\%~\mathrm{yr}^{-1}$), with all rates quoted relative to the long-term mean flux $\langle F_\gamma \rangle=7.8\times 10^{-10}\,\mathrm{erg}\,\mathrm{cm}^{-2}\,\mathrm{s}^{-1}$. Moreover, the flux of the LGF state is gradually approaching the stable HGF level at a rate of $+0.72 \pm 0.11\%~\mathrm{yr}^{-1}$. These results demonstrate that secular flux evolution in {\Star} operates largely independently of discrete state transitions, yet jointly with them drives the system toward a stable high-flux equilibrium.
\end{abstract}

\section{Introduction}           

{\Star} is one of the brightest $\gamma$-ray pulsars discovered by the {\fermi} Large Area Telescope (LAT) in blind searches \citep{Fermi-LAT:2009ihh, doi:10.1126/science.1175558}. It has a spin frequency $\nu \sim 3.8\,\mathrm{Hz}$, spin-down rate $|\dot{\nu}| \sim 8 \times 10^{-13}\,\mathrm{Hz\,s^{-1}}$, and characteristic age $\tau_c = 77\,\mathrm{kyr}$, with a double-peaked $\gamma$-ray pulse profile. An extended X-ray pulse peak is seen in {\xmm} observations \citep{Lin:2013boa}, and the pulsar may be associated with the $\sim$6.6-kyr-old supernova remnant G78.2+2.1 \citep{Trepl:2010pj}. Notably, {\Star} is radio-quiet \citep{Shaw:2023efh} yet exhibits state transitions---manifested as quasi-periodic, correlated variations in $\gamma$-ray flux $F_\gamma$, $\dot{\nu}$, and pulse shape---phenomena commonly observed in radio pulsars and thought to be linked to magnetospheric reconfigurations \citep{Kramer:2006ha, Hobbs:2009vh, Lyne:2010ad}, but exceptionally rare among $\gamma$-ray pulsars \citep{Fermi-LAT:2013wat}.

For most of the time, {\Star} resides in either a stable high $\gamma$-ray flux (HGF) state with low $|\dot{\nu}|$ and an emerging bridge component between the two pulse peaks, or a low $\gamma$-ray flux (LGF) state with elevated $|\dot{\nu}|$. However, during five observed state transitions---occurring approximately every 3 years---the pulsar transitions rapidly between states: HGF $\leftrightarrow$ LGF switches take place within $\sim$30-day, accompanied by $\sim$20\% variations in $F_\gamma$ and $\sim$4\% changes in $|\dot{\nu}|$ \citep{Fermi-LAT:2013wat, Ng:2016rpu, Takata:2020mdo, Fiori:2024nle, Wang:2023ioh}.

The apparent discontinuities in $\dot{\nu}$ during state transitions have led to glitch-inspired explanations, such as an increase in magnetic inclination angle caused by plates movement \citep{Ng:2016rpu}. Meanwhile, the quasi-periodic behavior of flux and spin-down rate has motivated interpretations invoking damped free precession of the neutron star \citep{Takata:2020mdo, Tong:2025ikn}. Multiwavelength observations further reveal a concurrent phase shift in the X-ray pulsations, hinting at a non-dipolar (e.g., quadrupolar) contribution to the magnetic field geometry \citep{Razzano:2023fic, Fiori:2024nle}. Yet a comprehensive picture of the long-term variability of {\Star} remains elusive.

It should be noted that, in the first HGF state, \citet{Fermi-LAT:2013wat} reported a $\sim$3-yr gradual flux increase prior to the transition and suggested a possible link to free precession \citep{Jones:2011jq}; however, no similar long-term intra-state evolution has since been reported. While \citet{Zhao:2017ssg} and \citet{Fiori:2024nle} modeled the $\sim$150-day post-transition flux recovery with a linear trend, they did not assume secular variation throughout the entire emission state. Leveraging the full 17-yr {\fermi}-LAT dataset and removing discrete state jumps, we reveal persistent secular variation in $F_\gamma$---correlated with $\dot{\nu}$ variability---\emph{within} individual emission states, beneath the quasi-periodic state transitions.

\section{Data reduction}\label{sec:data}
Our data are collected from August 5, 2008, to November 5, 2025, including LAT photons in the class of \texttt{P8R3\_SOURCE\_V3}, selected within a $15^{\circ}$-radius centered on the J2000 \textit{Chandra} position of {\Star} \citep{Weisskopf:2011da}. Data are in the energy range from $100\,\mathrm{MeV}$ to $300\,\mathrm{GeV}$, with zenith angles $z<90^\circ$.

For the flux analysis, we filtered the good time intervals by the  command \texttt{DATA\_QUAL>0\,\&\&\,LAT\_CONFIG==1}, restricted events to the Galactic point-source class (\texttt{evclass = 128}) and included both front- and back-converting events (\texttt{evtype = 3}). A binned likelihood analysis was performed, with the data divided into 35 logarithmically spaced energy bins (10 bins per decade) and spatial pixels of $0.1^{\circ} \times 0.1^{\circ}$.

For the timing analysis, we further confined the photon sample to a $1^{\circ}$-radius circular region around the source position and applied barycentric correction using the JPL~DE405 solar system ephemeris.

\section{Flux and timing analysis}\label{sec:method}
Binned likelihood analyses were conducted for each state and for consecutive 30-day intervals, via \textit{fermipy} v1.4.0 \citep{Wood:2017yyb} and \textit{Fermitools} v2.4.0, using the 4FGL-DR4 catalog \citep{Fermi-LAT:2022byn, Ballet:2023qzs} as the source model, where {\Star} was described by a power law with exponential cutoff:
\begin{equation}
   \frac{dN}{dE} =
\begin{cases}
N_0 \left(\frac{E}{E_0}\right)^{
  \gamma_0 
  - \frac{d}{2} \ln \frac{E}{E_0} 
  - \frac{d b}{6} \ln^2 \frac{E}{E_0} 
  - \frac{d b^2}{24} \ln^3 \frac{E}{E_0}
}, \\
\text{if } |b \ln \frac{E}{E_0}| < 10^{-2}; \\[2ex]

N_0 \left(\frac{E}{E_0}\right)^{\gamma_0 + d/b} 
  \exp \left( \frac{d}{b^2} \left(1 - \left(\frac{E}{E_0}\right)^b \right) \right), \\
\text{otherwise.}
\end{cases}
\end{equation}
The spectral parameters include the spectral slope $\gamma_0=\mathrm{d}\log(\mathrm{d}N/\mathrm{d}E)/\mathrm{d}\log E$ and the second derivative $-d=\mathrm{d}^2\log(\mathrm{d}N/\mathrm{d}E)/\mathrm{d}(\log E)^2$ at $E_0=1.8\,\mathrm{GeV}$. The exponent index $b$ is fixed to $0.39$.

We included all sources within a $25^{\circ} \times 25^{\circ}$ region centered on {\Star}, together with the Galactic and isotropic diffuse emission templates. The spectral parameters of {\Star} and the two diffuse components, and the normalizations of variable sources ($\texttt{Variability\_Index} > 18$ within $7^{\circ}$) and an extended source (G78.2$+$2.1) were freed in all fits. For each state, we identified sources with $\texttt{TS} > 20$ within $5^{\circ}$ and freed their normalizations. Specifically, the total numbers of free parameters from the first to the last state are: 30, 31, 31, 32, 28, 30, 27, and 30. This same set of freed parameters was then applied consistently to all 30-day intervals within the respective state.

Segmented timing analyses were performed with a second-order Taylor expansion model:
\begin{equation}
   \Phi - \Phi_0 = \nu (t - t_0) + \frac{1}{2} \dot{\nu} (t - t_0)^2,
\end{equation}
where $\Phi_0$, $\nu$, and $\dot{\nu}$ denote the phase, spin frequency, and frequency derivative at the reference epoch $t_0$, respectively. The length of each interval was determined adaptively based on the significance of the folded pulse profile, spanning from 40-day to 150-day.

Following timing ephemerides from \citet{Wang:2023ioh}, we folded pulse profiles and derived times-of-arrival (TOAs) from best frequencies. Once four or five TOAs were obtained, we fitted them with \texttt{TEMPO2} \citep{Hobbs:2006cd} to generate short-term ephemerides. Further details of the timing procedure can be found in \citet{Ge:2019dce}.

\begin{figure*}[]
   \centering
   \includegraphics[scale = 0.5]{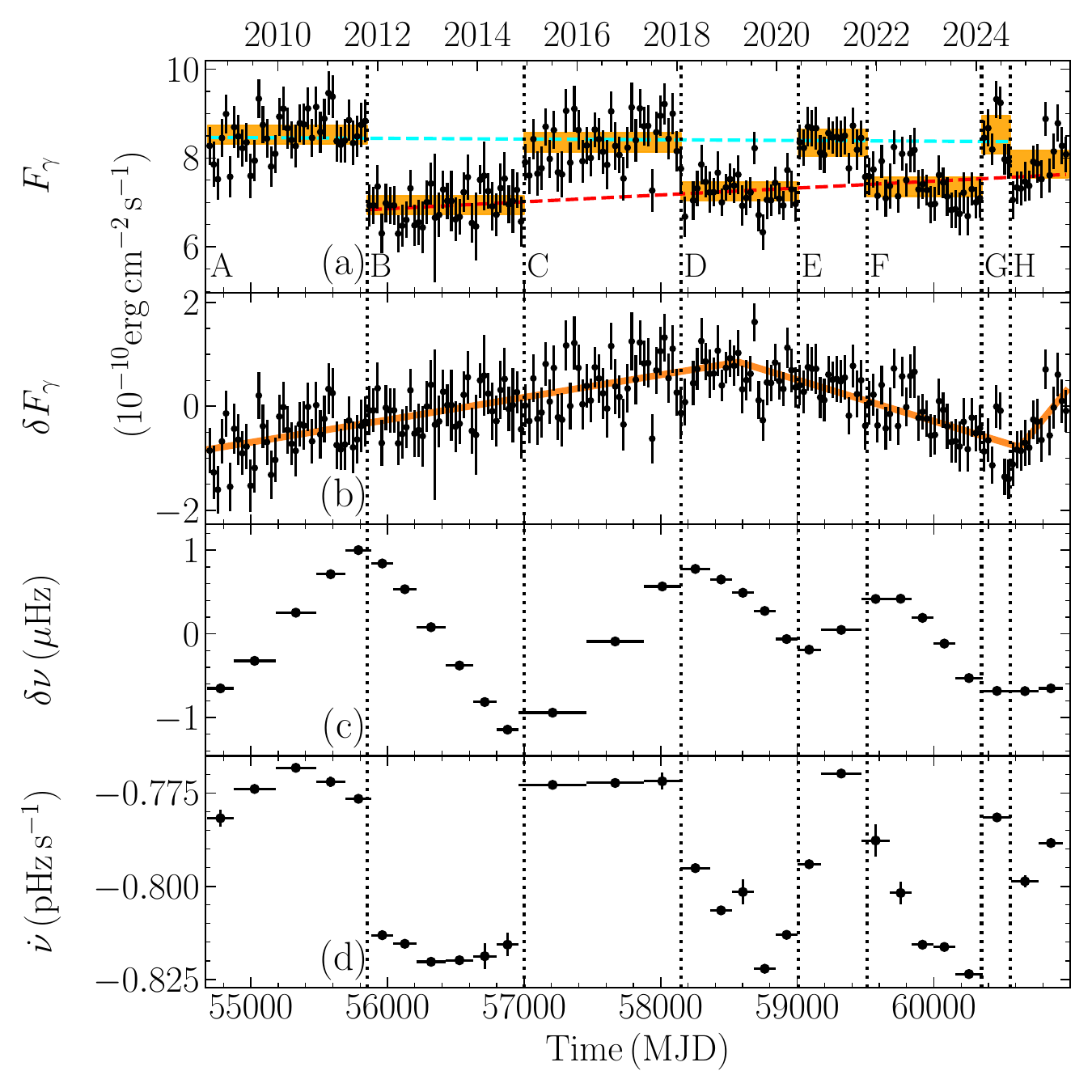}
   \caption{Energy flux and spin evolution of {\Star}.  
Panel (a): 30-day binned flux measurements with $1\sigma$ uncertainties (black points) and per-state fluxes with $3\sigma$ confidence intervals (orange bands). The cyan and red dashed lines are the linear fits to the fluxes in the HGF and LGF states respectively. 
Panel (b): Jump-corrected flux $\delta F_\gamma$, obtained by removing the mean offset and inter-state flux jumps; the orange piecewise linear curve shows the best fit to the secular evolution. 
Panels (c) and (d): Timing evolution in $\nu$ and $\dot{\nu}$; panel (c) displays the timing residuals $\delta \nu$ after subtracting the linear trend and mean offset. Vertical and horizontal error bars indicate $1\sigma$ uncertainties and ephemeris time spans, respectively. Vertical dotted lines mark the state transitions reported in \citet{Fiori:2024nle,Wang:2023ioh} and this work.}
   \label{fig:1}
\end{figure*}

\section{Results}\label{sec:results}

\begin{figure}[]
   \centering
   \includegraphics[scale = 0.5]{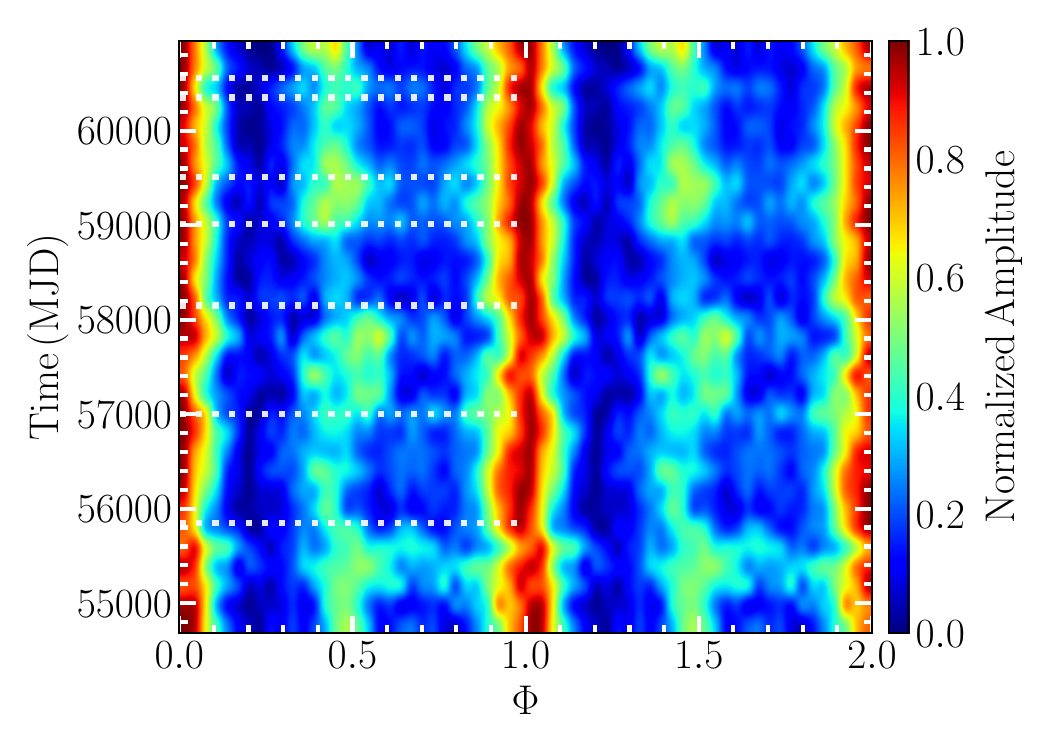}
   \caption{Two-dimensional histogram of the normalized pulse profile versus time, with phases and time each divided into 32 bins and 31 bins, smoothed via Gaussian interpolation. Horizontal dotted lines mark the state transitions. The main pulse, aligned to phase 0, is indicated in red.
}
   \label{fig:2}
\end{figure}

Relative to the mean flux $\langle F_\gamma \rangle$, the concurrent jumps in $F_\gamma$ and $\dot{\nu}$ (Figure~\ref{fig:1}), together with the associated pulse profile changes (Figure~\ref{fig:2}), confirm seven state transitions over the 17-yr baseline---two more than reported by \citet{Wang:2023ioh}. Following the convention of \citet{Fiori:2024nle}, we label the states sequentially with capital letters A through H and mark the transition epochs with dotted lines in both figures; detailed spectral and timing properties are provided in Table~\ref{table:spectrum} and the Zenodo repository \citep{liu_2025_18060028}\footnote{\url{https://doi.org/10.5281/zenodo.18060028}}.

To quantify intra-state evolution, we performed minimum $\chi^2$ linear fits to the 30-day binned flux in each state. The resulting flux evolution rates ($R_\mathrm{state}$), expressed relative to $\langle F_\gamma \rangle$ of each state, and inter-state flux jumps ($\Delta F_\gamma = F_{\gamma,\mathrm{end}}^{\mathrm{(prev)}} - F_{\gamma,\mathrm{start}}^{\mathrm{(next)}}$), derived from the best-fit parameters, are summarized in Table~\ref{table:spectrum}. In addition to the gradual rise in state~A, several other states exhibit non-zero $R_\mathrm{state}$ values, (e.g., $R_\mathrm{state} = +15\% \pm 4\%~\mathrm{yr}^{-1}$ in state~H), indicating that the long-term flux evolution observed in state~A is not isolated, but likely a persistent characteristic of {\Star}.

We further aligned adjacent states vertically by removing the flux jumps $\Delta F_\gamma$ and the mean offset, producing the jump-corrected light curve $\delta F_\gamma$ shown in Figure~\ref{fig:1}(b). A fit of the entire light curve $\delta F_\gamma$ with a constant flux model yields $\chi^2_\mathrm{red} = 2.21$ for 208 degrees of freedom, significantly exceeding the expected value of unity and confirming that $\delta F_\gamma$ deviates from a constant level. In contrast, a continuous piecewise linear model defined by 6 parameters provides an excellent description with $\chi^2_\mathrm{red} = 0.91$ (203 d.o.f). The best-fit solution comprises three distinct phases---a $\sim$10-yr rise, a $\sim$6-yr decline, and a recent $\sim$1-yr rapid rise—with phase-averaged rates of $R_\mathrm{phase} = +2.02^{+0.17}_{-0.15}\%~\mathrm{yr}^{-1}$, $-3.72^{+0.34}_{-0.47}\%~\mathrm{yr}^{-1}$, and $+14.9^{+6.4}_{-4.4}\%~\mathrm{yr}^{-1}$, and transition epochs at MJD~$58570^{+110}_{-93}$ and MJD~$60631^{+62}_{-65}$. Because the slope parameters and transition epochs are strongly correlated in the piecewise linear fit, we estimated their uncertainties using a residual bootstrap method; the asymmetric errors quoted correspond to the 16th and 84th percentiles of the bootstrap distributions.

Accompanying the intra-state flux evolution, $\dot{\nu}$ exhibits enhanced variability in states D--F, in contrast to the relatively stable spin-down rates in states A--C (Figure~\ref{fig:1}(d)). Notably, the total $\dot{\nu}$ variation in state~F is comparable in magnitude to the discrete jumps at the A--C transitions. Moreover, the elevated timing noise (MJD 58150--60350) coincides with the flux-decline phase, further supporting a physical link between secular flux evolution and spin-down behavior.

Additionally, despite intra-state flux variations and non-monotonic jump amplitudes, the long-term flux levels of the HGF states remain stable, whereas those of the LGF states show a gradual increase toward the HGF level---as illustrated by the orange confidence bands in Figure~\ref{fig:1}(a). Linear fits to all 30-day binned fluxes of HGF and LGF (cyan and red dashed lines) yield evolution rates of $R_\mathrm{HGF} = -0.08 \pm 0.11\%~\mathrm{yr}^{-1}$ and $R_\mathrm{LGF} = +0.72 \pm 0.11\%~\mathrm{yr}^{-1}$, respectively. This differential evolution---arising from the interplay of discrete state transitions and secular trends---suggests that {\Star} is undergoing a dissipative relaxation toward a stable HGF equilibrium state. Extrapolation indicates that the HGF and LGF flux levels will converge near MJD~65250 (2037).

\begin{deluxetable*}{lccccccccc}[htb]\label{table:spectrum}
\tablecaption{Spectral parameters and the flux variations of {\Star} in different states.}
\tablehead{
\colhead{Parameter} & 
\colhead{A} & 
\colhead{B} & 
\colhead{C} & 
\colhead{D} & 
\colhead{E} & 
\colhead{F} & 
\colhead{G} & 
\colhead{H}
}
\startdata
Start & 54683 & 55850 & 57000 & 58150 & 59010 & 59510 & 60350 & 60560 \\
Stop & 55850 & 57000 & 58150 & 59010 & 59510 & 60350 & 60560 & 60984 \\
$F_\gamma$\tablenotemark{\footnotesize a} & 8.52(7) & 6.94(7) & 8.35(7) & 7.26(7) & 8.3(1) & 7.36(7) & 8.5(1) & 7.9(1) \\
$F_{\gamma, \mathrm{pho}}$\tablenotemark{\footnotesize b} & 1.09(1) & 0.95(1) & 1.08(1) & 1.00(1) & 1.11(2) & 1.01(1) & 1.15(3) & 1.04(2) \\
$\gamma_0$ & -2.49(1) & -2.61(2) & -2.52(1) & -2.61(2) & -2.49(2) & -2.59(2) & -2.52(3) & -2.54(2) \\
$d$ & 0.75(2) & 0.81(2) & 0.77(2) & 0.80(2) & 0.69(2) & 0.77(2) & 0.72(3) & 0.77(3) \\
$R_\mathrm{state}$\tablenotemark{\footnotesize c} & $2.6\pm1.0$ & $1.7\pm1.0$ & $2.3\pm1.0$ & $-0.9\pm1.5$ & $-3.1\pm3.2$ & $-4.4\pm1.5$ & $-11\pm11$ & $15\pm4$ \\
$\Delta F_\gamma$\tablenotemark{\footnotesize d} &  & -2.1 & 0.88 & -1.3 & 1.4 & -0.43 & 1.8 & -1.2 \\
\enddata
\setlength{\leftskip}{5.5em}
\tablecomments{The numbers in the parentheses denote the 1$\sigma$ errors in the last digit.}\vspace{-7pt}
\tablenotetext{a}{Energy flux in units of $10^{-10}\,\mathrm{erg}\,\mathrm{cm}^{-2}\,\mathrm{s}^{-1}$.}\vspace{-7pt}
\tablenotetext{b}{Photon flux in units of $10^{-6}\,\mathrm{photon}\,\mathrm{cm}^{-2}\,\mathrm{s}^{-1}$.}\vspace{-7pt}
\tablenotetext{c}{Relative changing rate of flux in units of percentage per year.}\vspace{-7pt}
\tablenotetext{d}{Jump of energy flux from the previous state in units of $10^{-10}\,\mathrm{erg}\,\mathrm{cm}^{-2}\,\mathrm{s}^{-1}$.}\vspace{-25pt}

\end{deluxetable*}

\section{Discussion and summary}\label{sec:dis_and_sum}
Beyond the discrete flux jumps at state transitions, the intra-state variability and three-phase secular evolution uncovered here add a new dimension to the long-term behavior of {\Star}. This suggests that the magnetic inclination angle or polar-cap magnetospheric configuration \citep{Ng:2016rpu,Zhao:2017ssg,Fiori:2024nle} may undergo changes---both abrupt and gradual. We caution, however, that our analysis treats state transitions as instantaneous and discontinuous events; future work will be needed to assess whether they possess short-term aftereffects that could influence the inferred secular evolution.

In its flux evolution, {\Star} displays two secular features: a quasi-oscillatory variation of $\delta F_\gamma$ and a systematic convergence of LGF toward the stable HGF level. These characteristics share qualitative similarities with the oscillatory and dissipative aspects of damping precession \citep{Tong:2025ikn}. Yet, the increase in $|R_\mathrm{phase}|$ over time and the stability of the HGF state are not naturally explained within that framework, suggesting that additional complexities in the magnetospheric or interior dynamics of {\Star} may be at play. Using the upper limit on $\Delta \nu$ and the measured $\Delta \dot{\nu}$ at the state transitions, \citet{Zhao:2017ssg} found that the relative changes $\Delta \nu / \nu$ and $\Delta \dot{\nu} / \dot{\nu}$ in {\Star} are consistent with the range observed in glitches for young pulsars. Nevertheless, we find that when $R_\mathrm{phase}<0$, the variability of $\dot{\nu}$ within each state is enhanced, which further increases the difficulty of distinguishing potential glitch signatures at the D--F state transitions.

It is worth noting that \citet{Razzano:2023fic} reported a $\sim$0.25 phase lag in the X-ray pulse profile of {\Star} between two {\xmm} observations separated by 3.7~yr, spanning the B--C state transition epoch. They interpreted this shift as a reconfiguration of the coupling between the near-surface quadrupole magnetic field and the large-scale dipole component. However, apart from this isolated shift, no significant spectral or morphological variability has been detected in the X-ray emission \citep{Wang:2018vez}, which possesses the largest known $\gamma$-ray-to-X-ray flux ratio ($\sim 6.5 \times 10^4$) \citep{Hui:2016cim}. This decoupling may reflect a special magnetospheric \citep{Fiori:2024nle}, or it could be just limited by the precision of observations.

Both the Crab and Vela pulsars exhibit glitches but no state transitions; an observable long-term decline in X-ray flux is seen in Crab, correlated with its decreasing spin-down luminosity \citep{Yan:2018knt,Wang:2023ioh}. In contrast, for {\Star}, the relationship between flux and spin evolution is more complex, as detailed in \hyperref[appen]{Appendix}. State transitions are more common in radio pulsars and are classified as one type of timing noise, attributed to magnetospheric reconfigurations \citep{Kramer:2006ha, Lyne:2010ad}. A comparable case is PSR~J1124$-$5916---a young pulsar with radio and $\gamma$-ray pulsations that exhibits $\dot{\nu}$-state transitions and, owing to its low timing noise, has yielded a precise braking index measurement \citep{Ge:2020jrj}. However, unlike {\Star}, its state transitions show no detectable changes in $F_\gamma$ or pulse profile, and its spin evolution is likely dominated by magnetic dipole radiation and pulsar wind. Similarly, PSR~B0943$+$10 shows correlated radio and X-ray state switching---accompanied by profile transformations---yet displays no significant change in spin properties \citep{Hermsen:2013zs}.

To advance our understanding of state transitions, secular evolution, and their relationship with spin-down behavior, coordinated multiwavelength monitoring---especially in X-rays---is essential. The \emph{Einstein Probe} (\emph{EP}; \citealt{Yuan:2022fpj, Yuan:2025cbh}), and the upcoming missions \emph{enhanced X-ray Timing and Polarimetry mission} (\emph{eXTP}; \citealt{Zhang:2025iae, Ge:2025upe}) will enable high-precision pulse profile and polarization measurements over the next decade. Sustained, simultaneous coverage by {\fermi}-LAT and these X-ray facilities is key to unraveling the physical linkage between $\gamma$-ray and X-ray emission processes in pulsars.

\begin{acknowledgements}
This work is supported by the China's Space Origins Exploration Program and the National
Natural Science Foundation of China (Grant Nos. 12373051, 12033001, 12273043, 12473039). We acknowledge the use of the public data from the \emph{Fermi}-LAT data archive. We thank Yu-Long Yan, Qian Zhang, Pei-Xin Zhu and Jia-Chen Xie for helpful discussions on pulsar precession and glitch phenomena.

\end{acknowledgements}

\appendix
\phantomsection \label{appen}
\twocolumngrid
To investigate a potential connection between $\delta F_\gamma$ and spin behavior, we constructed a time series of $\delta |\dot{\nu}|$ following an analogous procedure to that used for $\delta F_\gamma$, aiming to remove the influence of discrete state transitions. However, due to the limited number of timing epochs, the difference in $|\dot{\nu}|$ between the two states was simply taken as the difference between two adjacent measurements. To enable a direct comparison, we also resampled the original $\delta F_\gamma$ light curve using weighted averages over time intervals matched to those of the $\dot{\nu}$, resulting in 29 data points for both $\delta F_\gamma$ and $\delta |\dot{\nu}|$.

\begin{figure}[]
   \centering
   \includegraphics[scale = 0.67]{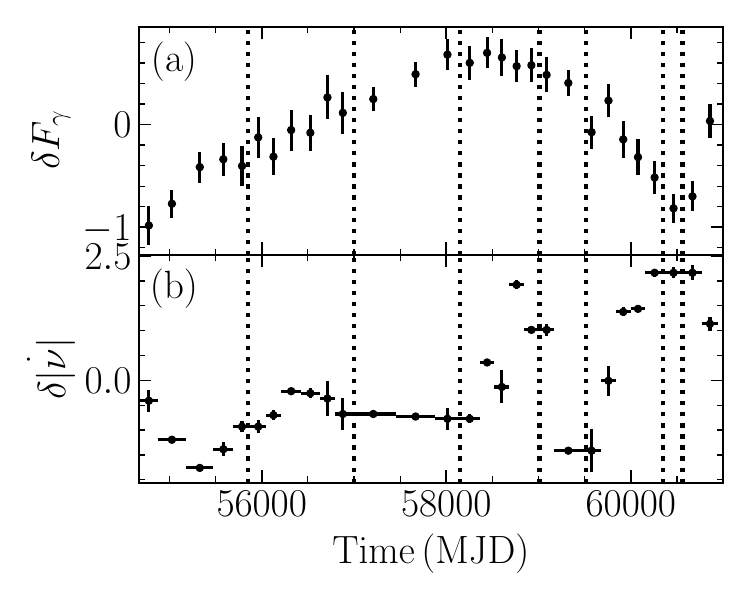}
   \caption{
Panel (a): Resampled $\delta F_\gamma$ from weighted averages, in units of $10^{-10}\,\mathrm{erg}\,\mathrm{cm}^{-2}\,\mathrm{s}^{-1}$.
Panel (b): Jump-corrected $\delta |\dot{\nu}|$ in units of $10^{-14}\,\mathrm{Hz}\,\mathrm{s}^{-1}$, obtained by subtracting the inter-state timing jumps and mean offset. Dotted lines mark the state transitions.
}
   \label{fig:3}
\end{figure}

As shown in Figure~\ref{fig:3}, a visual inspection reveals little correspondence between $\delta F_\gamma$ and $\delta |\dot{\nu}|$. Quantitatively, their Pearson correlation coefficient is $-0.093$ with a $p$-value of $0.631$, indicating no significant linear correlation. In contrast, when the same resampling is applied to the $F_\gamma$, we find a strong anti-correlation with $|\dot{\nu}|$: the Pearson coefficient is $-0.91$ ($p$-value $\sim 10^{-12}$). This stark difference highlights the complexity of the relationship between $\gamma$-ray emission and spin-down evolution in {\Star}.

\bibliographystyle{aasjournal}
\bibliography{main}

\end{document}